# Near-infrared scintillation of liquid argon

**T. Alexander,**[a] **C.O. Escobar,**[a,b,1] **W.H. Lippincott**[a] **and P. Rubinov**[a]

[a]*Fermi National Accelerator Laboratory,*
 *Kirk Rd and Pine St., Batavia, IL 60510, U.S.A.*

[b]*Instituto de Física da Universidade Estadual de Campinas,*
 *rua Sergio Buarque de Holanda 777, 1308-859, Campinas, SP, Brazil*

 *E-mail:* escobar@fnal.gov

Abstract: Since the 1970s it has been known that noble gases scintillate in the near infrared (NIR) region of the spectrum (0.7 µm < λ < 1.5 µm). More controversial has been the question of the NIR light yield for condensed noble gases. We first present the motivation for using the NIR scintillation in liquid argon detectors, then briefly review early as well as more recent efforts and finally show encouraging preliminary results of a test performed at Fermilab.

Keywords: Scintillators, scintillation and light emission processes (solid, gas and liquid scintillators); Noble liquid detectors (scintillation, ionization, double-phase)

---

[1]Corresponding author.

# Contents

## 1 Introduction

**Motivation for using NIR light.** It is well-known that the detection of both charge and light signals is a desirable feature in many experiments using liquid noble gases (LNG) [1], ranging from dark matter (DM) searches [2] to large liquid argon (LAr) time projection chambers (TPC) used in neutrino physics [3]. For the latter the scintillation signal is important in providing the start time $t_0$ for non-accelerator related physics, while for DM experiments the detection of the light signal provides background rejection by separating nuclear recoils from electronic ones.

So far the vast majority of the current or planned experiments focus on the detection of the vacuum-ultraviolet (VUV) scintillation light from noble gases either in gaseous or condensed states with wavelengths as short as 78 nm for liquid neon up to 175 nm in liquid xenon. The detection of the VUV light presents many experimental challenges starting with the lack of availability of reasonably priced (cost per unit sensitive area) cryogenic VUV sensitive photodetectors. Most argon detectors use wavelength shifters (WLS) such as tetra phenyl butadiene (TPB) [4], for example in the light-guide bars designed for the DUNE experiment [5] or directly coating a visible light sensitive photodetector as for the ArDM experiment [6]. The use of TPB raises concerns about long term stability and extreme care in the storage and handling of coated surfaces due to the degradation of TPB when exposed to UV light or by its oxidation in air or hydration when kept in non-dry atmospheres [7]. Another challenge, significant for very large TPC with long drift distances, is the presence of Rayleigh scattering which introduces pernicious effects such as a two to three-fold decrease in the number of collected photons coming from distances at or beyond 2.5 m. Poor reflectivity of VUV light from most materials make reflective surfaces, unless coated with TPB, almost useless and finally, Rayleigh scattering smears the time resolution [8]. These negative effects introduced by Rayleigh scattering are brought to the fore if a recent estimate of the Rayleigh scattering length [9] placing it at 55cm is taken into account. Lastly it should be mentioned that the recently observed delayed light emission from TPB makes the detection of the triplet light harder, since it occurs on a longer time scale than the singlet for LAr [10]. All of the above difficulties cease to exist if LNGs have a significant emission in the NIR and that light is used either as a replacement or in addition to the VUV signal.

One further advantage of using the NIR scintillation as the light signal in large LAr TPCs is the possibility for doping LAr with a photosensitive chemical such as tetra-methyl-germanium (TMG) to help recover the charge lost by recombination, as proposed by the ICARUS collaboration [11]. TMG due to its large photo-absorption cross-section significantly reduces the amount of useful VUV light while any NIR light would travel without any attenuation, enhanced charge and light collection would go hand in hand, violating the complementarity [1] between charge and light signals.

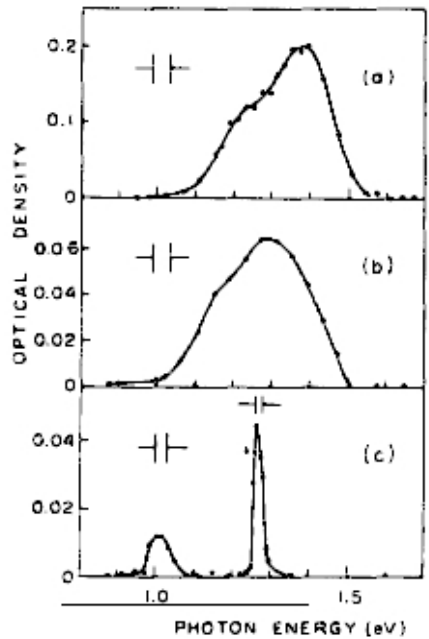

**Figure 1**. Spectra of transient absorption of Ar. (a) solid; (b) liquid and (c) gas. From ref. [14].

## 2 Brief historical review

**Early results.** Near-infrared emission from noble gases has been known since the late 1940s with emission lines around 1,300 nm [12]. Using the technique of transient optical absorption spectroscopy via electron excitation, NIR transitions were observed in noble gases both in the gaseous state [13] and in condensed form [14]. The similarity of the spectral features in gas and in liquid or solid in the particular case of argon, as seen in figure 1, from reference [14], has led to the interpretation that these features are due to the formation of self-trapped excitons.

In all three phases the location of the peak is around 1.27 eV which corresponds to a wavelength of 976 nm. In the gas phase one further peak is observed around 1 eV (1,240 nm).

The interest on the NIR emission from noble gases resurfaced with work of Lindblom and Solin [15] who observed atomic lines in the NIR under excitation by a low-energy (3–5 MeV) proton beam. This was followed by a systematic investigation of NIR emission from liquid and gaseous Ar and Xe by Bressi, Carugno and co-workers [16], with the goal of developing new particle detectors. Their results showed that both xenon and argon scintillate in the NIR, but the results were inconclusive for the liquid state of both noble gases, as the light yield was poorly estimated [17].

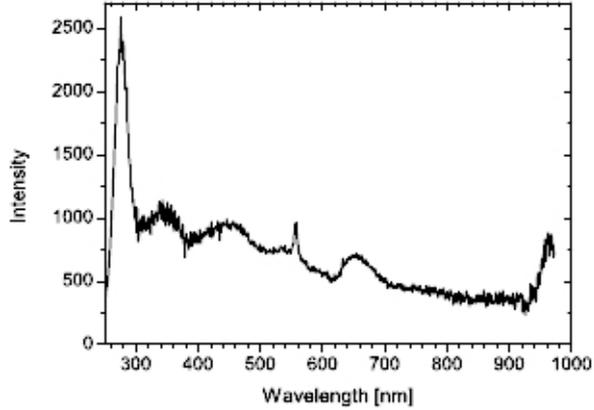

**Figure 2**. Results from the Munich group [20]: emission spectra of LAr at 85K. The peak at 557 nm is identified as due to oxygen impurity. For our purposes notice the spectral feature at 970 nm (close to the cut-off wavelength of their spectrometer).

**Recent results.** In the last five years, two groups have pursued the investigation of NIR scintillation in LAr, one based in Novosibirsk, Russia [18] and the other at the Technical University of Munich, Germany [19]. Both groups use table-top setups, with volumes of a few cubic centimeters of LAr excited by very intense, low-energy beams: 12 keV electrons (pulsed or continuous) for the Munich experiments and pulsed X-rays with energies between 30 and 40 keV for the Novosibirsk setup. We shall now briefly describe the techniques used by each group and summarize their results.

The Munich group studied the scintillation of LAr over a wide range of wavelengths, from the VUV to the NIR, using a monochromator and a VUV sensitive photomultiplier for their study of the second and third continua (up to 320 nm) and a spectrometer to study the region from 250 nm to 1000 nm. No absolute light yield is provided but several spectral features were associated with traces impurities. Figure 2 shows a light spectrum from one of their earliest publications [20].

Later, we will return to the more recent results from the Munich group but now we would like to summarize the results obtained by Bondar et al. [18, 21, 22], which are revelant for our own preliminary test, which we will report in the next section.

The Novosibirsk group employed silicon photo multipliers (SiPM) with sensitivity peaked around 600 nm, extending to 1,000 nm. As they do not use a spectrometer, in contrast to the Munich group, they obtain an integrated scintillation yield by convoluting the photon detection efficiency of their SiPMs with the spectrum obtained by the Munich group and the geometrical acceptance of their apparatus (shown in figure 3), obtained by Monte Carlo simulation. The normalization of the scintillation spectrum is thus obtained by matching the number of counted photoelectrons in the SiPMs.

Bondar et al. obtained a light yield of $1.7 \times 10^4$ photons/MeV in gaseous Ar in the range 690–1,000 nm and of $5.1 \times 10^2$ photons/MeV in LAr in the range 400–1,000 nm. It is important to emphasize that Bondar et al. use in their analysis the unnormalized spectrum obtained by Heindl et al. [19, 20] as this leads us to review a recent result from the Munich group that points out to the difficulties intrinsic to this kind of measurement and its interpretation.

More recently, the Munich group has published new results of their on-going studies of the scintillation of pure liquid argon and liquid Ar-Xe mixtures, stating that they no longer see any relevant emission in pure liquid argon in the range 500 to 3,000 nm [23]. They therefore revise their previous result ([19, 20]) saying that they now believe the observed signal was an "artifact due to the normalization of the spectrum with the response function of the spectrometer used for that spectral range".

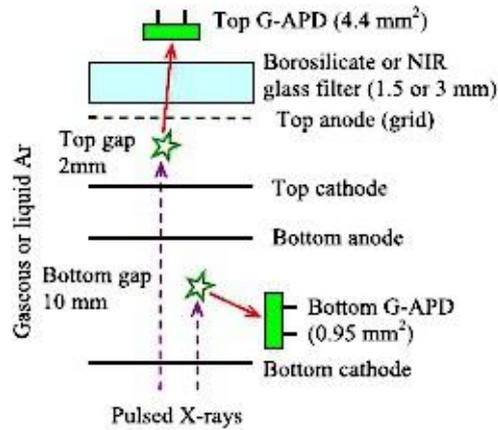

**Figure 3**. Lay-out of the experimental setup used by Bondar et al. The bottom SiPM (G-APD) is used without optical filters. From [18].

**Lessons learned.** It seems to us that a vigorous, systematic, experimental program investigating the NIR scintillation of LNGs is needed, one that would use larger volumes of LNGs and aiming at

- Rigorous control of purity
- Obtaining spectral information
- Determining the time structure of the NIR emission
- Determining the light yield in the NIR for different sources of ionizing radiation
- Exciting the media with radiation more akin to the radiation that will be encountered in real applications such as in DM and/or neutrino physics detectors.

With these goals in view we have started a research program at Fermilab and as a first step made a preliminary test of NIR light emission with encouraging results as reported in the next section.

## 3 Results from a preliminary test

Before embarking on a full R&D program for the investigation of the NIR scintillation of LNGs, we performed a test using a cryostat that was used before by the Scene collaboration for measuring the light yield of low energy nuclear recoils in LAr [24]. To detect the scintillation light we chose the same SiPMs used by Bondar et al. [18], model 149-35 manufactured by CPTA with sensitivity peaked around 600 nm and extending to almost 1,000 nm (figure 4) [25]. Seven SiPMs were placed on a plastic ring of 2 inches diameter with a red LED occupying the 8$^{th}$ slot on the ring. The ring was immersed into the liquid argon and placed some 5 inches below the liquid-gas interface in the inner cryostat vessel. A 1μCi $^{22}$Na positron source was used to produce two positron annihilation gamma rays (0.511 MeV) with opposite momenta. We tag the backward-going gamma ray using a liquid scintillator counter (EJ301 liquid scintillator from Eljen) read-out by a photomultiplier tube. The schematic lay-out of the setup is shown in figure 5.

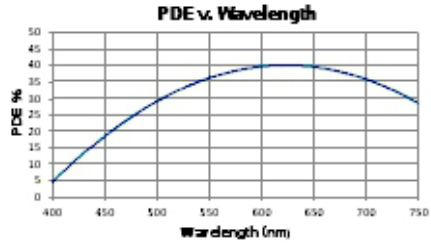

**Figure 4**. Photo detection efficiency of the CPTA 149-35 SiPM [25].

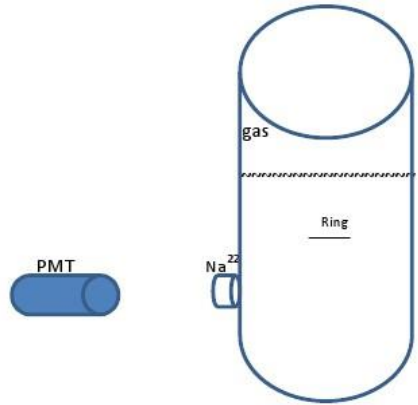

**Figure 5**. Configuration of the experiment showing tagging detector (PMT).

The readout chain for the experiment consists entirely of warm electronics positioned outside the cryostat. The connection from the cryostat is carried on 50 ohm RG58 coaxial cables, with the bias and signal on the inner conductor and reference ground on the shield. The main amplifier used, the Photonique model AMP-0611 was packaged in a custom NIM module, supporting up to 8 amplifiers, bias distribution and filter circuits as well as bias adjusting trim pots. The bias voltage was generated by a Keithley 2400 SourceMeter. The power from the amplifiers was supplied by a standard NIM crate. The outputs from the NIM module were routed to a Tektronix DPO3054 digital oscilloscope which was used as the main digitizer. The readout was accomplished via a small module within Excel using the USB interface of the scope. The scope was used to trigger on the signals and could be used in single channel or "Logic OR" mode for up to 4 channels. In addition, the external trigger output of the DPO3054 was also routed to a HP53131A counter timer unit (read out via GPIB) to allow for measurement of signal rates.

With the above configuration the only degrees of freedom available for introducing variation in the setup were the height of the source and counter with respect to the ring with the SiPMs (or the gas-liquid interface) and the distance between the radioactive source and the scintillation counter which defines the solid angle covered by the annihilation gamma-rays. The rates are expected to be low due to the low source intensity, small geometrical acceptance and small energy loss of the gammas in the LAr (mostly to Compton electrons). We ran the SiPMs at a bias voltage of 44 V. To avoid excessive noise that arose when using multiple SiPMs in coincidence, we trigger on the best SiPM at a fraction of a single p.e., looking for the EJ301 PMT pulse inside a time window that spans negative (PMT fires before the trigger) and positive times (PMT signal after the SiPM). If the SiPM is triggering on scintillation light we expect to have more PMT signals occurring earlier (negative times) than later (positive times). Figure 6 shows a typical collection of scope traces from the SiPM and the scintillation counter (the time axis has the origin translated so as to have only positive times). A cut is placed on the pulse height of the scintillator PMT traces so as to select the 0.511 MeV gamma, eliminating the 1.275 MeV gamma that also comes from the decay of $^{22}$Na. As a consistency check we verified that the latter signal is evenly spread in early and late time regions, as it should since it is not correlated with the scintillation signal. These results confirm that ionizing radiation in LAr produces light with wavelength between 600 and 1000 nm. The light yield in this region has not been determined due to the low geometrical acceptance of our experimental setup and the small energy loss in the corresponding volume of LAr.

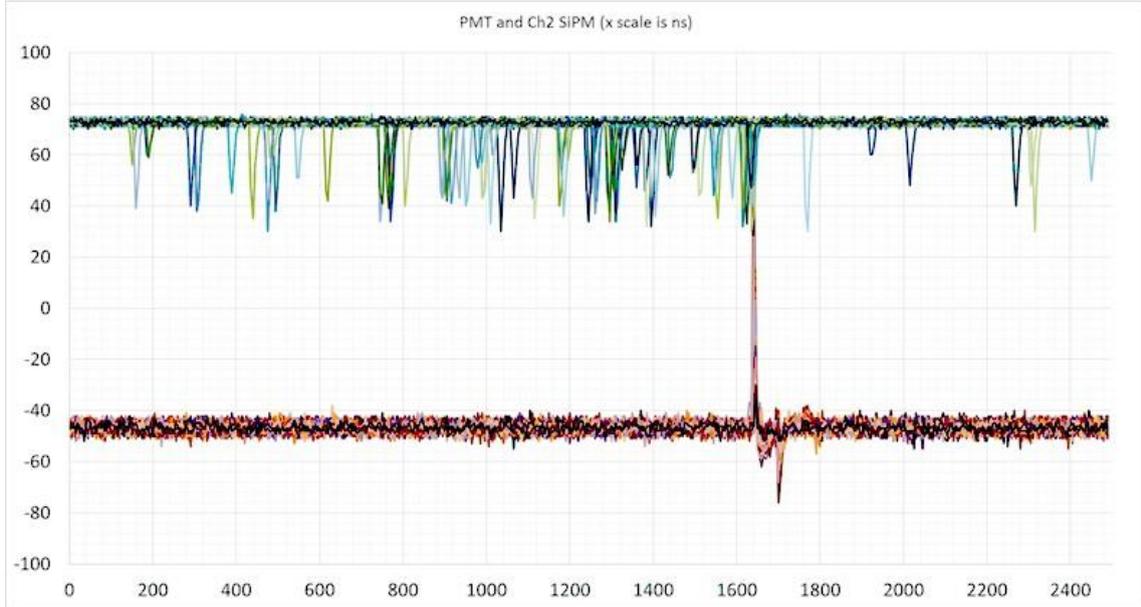

**Figure 6**. Scope traces from the SiPM (bottom) and PMT (top) showing predominance of early time PMT signals as expected if the SiPM is triggering on scintillation light.

Figure 7 shows a typical time distribution from the PMT signals, which we fit with a double exponential. The results show a fast component with a time constant of less than 200 ns and a slow one with 2.72 μs. This result is consistent with the time distribution obtained by Bondar et al. [21, 22] who also obtained short and slow components, with the latter remaining without explanation regarding its origin.

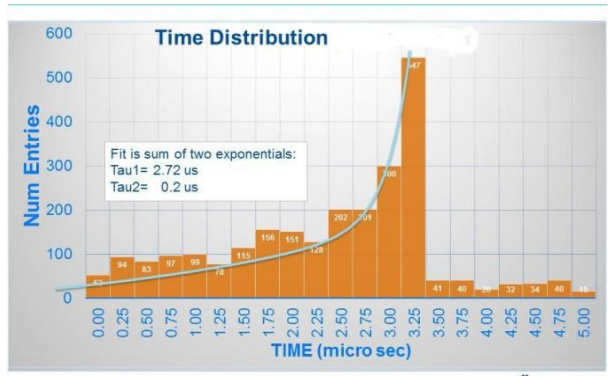

**Figure 7**. Time distribution of the scintillation counter signal. Notice the presence of a fast and a slow component.

## 4  Conclusions

NIR scintillation from liquid noble gases is a promising alternative to VUV light for determining the start time of an event ($t_0$), helping with particle identification and the reconstruction of very low-energy events, which is important for the detection of supernova neutrinos and search for proton decay [26]. The use of NIR light would by-pass problems and difficulties associated with the use of VUV scintillation and opens the door for simultaneously improving charge collection in Lar TPC's through doping, avoiding the quenching of the scintillation light. The simultaneous detection of the VUV and NIR signals in DM experiments could help separate nuclear recoils from the electromagnetic background if, as expected on theoretical grounds, the NIR light is more abundant for denser ionization tracks.

In order to fulfil these goals we plan to continue investigating the NIR scintillation in LNGs addressing the challenges outlined in section 2.3.

## Acknowledgments

We would like to thank Michael Reid for a GEANT4 simulation that helped us check the soundness of our results. The technical staff at the Fermilab Proton Assembly Building is thanked for their support. Fermilab is Operated by Fermi Research Alliance, LLC under Contract No. De-AC02-07CH11359 with the United States Department of Energy.